\documentclass[aps,twocolumn,showpacs,showkeys,nofootinbib]{revtex4-1}
\usepackage{epsfig}
\usepackage{amsmath}
\usepackage{amsfonts}
\usepackage{amssymb}
\usepackage{graphicx}
\usepackage{colordvi}
\usepackage{color}
\usepackage{dcolumn}
\usepackage{multirow}
\usepackage{hyperref}
\usepackage{epstopdf}
\usepackage{bm}
\usepackage{times}
\hypersetup{
  colorlinks=true,        
  linkcolor=blue,         
  citecolor=magenta,      
}

\begin{document}

\title{Cosmological analysis of a DGP stable model with $H(z)$ observations a revision}

\author{Gabriela B\'arcenas-Enr\'iquez}
\email{gabrielabarcenas94@gmail.com}
\affiliation {Unidad Acad\'emica de F\'isica, Universidad Aut\'onoma de Zacatecas, Calzada Solidaridad esquina con Paseo a la Bufa S/N C.P. 98060,Zacatecas, M\'exico}

\author{Celia Escamilla-Rivera}
\email{cescamilla@mctp.mx}
\affiliation {Mesoamerican Centre for Theoretical Physics,\\ Universidad Aut\'onoma de Chiapas.
Ciudad Universitaria, Carretera Zapata Km. 4, Real del Bosque (Ter\'an), 29040, Tuxtla Guti\'errez, Chiapas, M\'exico.}

\author{Miguel A. Garc\'ia-Aspeitia}
\email{aspeitia@fisica.uaz.edu.mx}
\affiliation {Unidad Acad\'emica de F\'isica, Universidad Aut\'onoma de Zacatecas, Calzada Solidaridad esquina con Paseo a la Bufa S/N C.P. 98060,Zacatecas, M\'exico. Consejo Nacional de Ciencia y Tecnolog\'ia, Av. Insurgentes Sur 1582. 
Colonia Cr\'edito Constructor, Del. Benito Ju\'arez C.P. 03940, Ciudad de M\'exico, M\'exico}


\begin{abstract}
In this paper, we will present a Dvali-Gabadadze-Porrati stable model in order to perform an observational 
test using $H(z)$ data and radial BAO scale in the galaxy distribution. In this vein, we study the tension 
between constraints on the cosmological constant $\Lambda$ and the crossover scale $r_c$, which is associated 
with the DGP model. Our results show that observations do not favor the DGP stable model as a possible 
candidate to fit the observations of the late cosmic acceleration.
\end{abstract}


        
\maketitle

\section{Introduction}

One of the central challenges of modern cosmology is still to shed light on the physical mechanism behind the cosmic acceleration. Current measurements have already sharply improved constraints on this phenomena. Several observations like Supernovas SNeIa \cite{Riess:1998cb}, Cosmic Microwave Background Radiation (CMBR) \cite{Spergel:2003cb}, Baryonic Acoustic Oscillations (BAO) \cite{Eisenstein:2005su}, among others \cite{Tegmark:2003ud,Jain:2003tba,Magana,Riess:2016jrr}, has been useful to constraints the cosmological parameters that define a specific model. Future observations are expected to do much better, especially for models that allow a time-evolving Equation of State (EoS).

One of the main candidate to explain this cosmic acceleration is Dark Energy (DE). This component also features baryonic matter, dark matter and radiation. The advantage of DE is that relax some tensions in the cosmological parameters measurements, which can explain in particular the fact the geometry of the universe is consistent with the flatness predicted by inflation. Despite the large observational progress in measuring DE properties, not fundamental insights into the physics behind this dark sector has been solved. Even thought, while the statistical error have shrunk dramatically, current constraints are still roughly consistent with 68.3\% \cite{Ade:2015xua} current energy budget with an EoS ratio $\omega\approx-1$. This had led to the idea in where a Cosmological Constant (CC) $\Lambda$ can explain the cosmic acceleration. Also, in agreement with the described observations, the $\Lambda$CDM or concordance model has the advantage to provide an accelerated behavior driven by $\Lambda$ and filled with Cold Dark Matter (CDM).

Despite its simplicity, there are fundamental problems if we assume that CC is related with the quantum vacuum fluctuations. Some theoretical efforts point out to a value of density energy $\sim120$ orders of magnitude of difference with the observational value or at least it is expected a strictly vanishing value under protective symmetry \cite{Weinberg}. For this reason, some research has turned to find new alternatives as: quintessence \cite{Qreferences}, phantom fields \cite{Caldwell:2003vq}, Chaplygin models \cite{Bento:2002ps}, brane models \cite{Aghababaie:2003wz}, just to mention a few. 

Between the plethora of models, one of the most interesting alternatives comes from a brane model based in the idea of Dvali, Gabadadze and Porrati (DGP), where it is assumed a $5$D Minkowski space time, within a $4$D Minkowski brane embedded  \cite{dgpmodels}. The region of transition between the four and five dimensional manifold is encoded in the crossover scale parameter $r_c$, which is a function of the five and four Planck masses. It is interesting to notice that this scenario allows to mimic the universe acceleration as a transition between the dimensions of spacetime mimicking the CC with the crossover region parameter. A natural extension of DGP models can be performed when the brane is generalized 
by using a Friedmann-Lemaitre-Robertson-Walker (FLRW) metric. Therefore, this model offers an attractive explanation for the accelerated expansion of the universe without to invoke a dark energy component. From the DGP background evolution emerge two solution branches depending the choice of the sign: the self-accelerated branch (which corresponds to a negative sign) and the normal or stable branch (which correspond to a positive sign). In the first branch there is a late cosmic acceleration without the presence of DE. However, this branch is ruled out by supernovae data \cite{fang}. The normal/stable branch has the property of introducing a $\Lambda$ over the evolution and fixed a bidimensional model with two cosmological parameters: $\Omega_{rc}$ and $\Omega_{\Lambda}$. This latter characteristic allow us to perform a directly astrophysical test using $H(z)$ data set and Planck analysis \cite{Ade:2015xua} which can set constraints on the cosmological parameters of the stable DGP model. However, the impact of these cosmological parameters will became looser (stronger) depending of the weakness (strength) of the fifth force. Interesting results related to these cases are reported in \cite{Aguilera:2013tmp,Barreira:2016ovx}, studying a IR cutoff or the growth rate of structure or in \cite{Raccanelli} it was studied tests of gravity using large-scale redshift-space distortions.

In this paper, we will work with the stable DGP model for three different pipelines in where we can control the strength of the $r_c$ parameter and set the constraints over this parameter using direct $H(z)$ measurements: the Cosmic Chronometers (Cosmic-C) and the radial BAO scale in the galaxy distribution. In \cite{Wan:2007pm} was study a DGP universe using these observations, however the variation of the curvature in this analysis shows a DGP model with best fits that correspond to a/an closed/open universe using a WMAP prior.

This paper is organised as follows. In Sec. 2 we will present an overview of the equations related to the DGP background cosmology. In Sec. 3 we describe the astrophysical samples for $H(z)$. In Sec. 4 we present the constraints over the DGP cosmological parameters of our interest. In Sec. 5 with set a discussion of the results obtained.

\section{DGP cosmological background}
The DGP model \cite{Dvali:2000hr} suggest an universe on a brane which is embedded in a 5D Minkowski space-time with a infinite extra dimension.
This model give us two important reasons to consider it. First, it describe a 4D Newtonian gravity on the brane at short distances whereas on the bulk the gravity shows as 5D. Second, the short distances are fixed by a crossover scale $r_c$ denoted by $r_c \equiv M^2_P/2M^3$, where $M_P$ and $M$ are the five and four Planck masses, respectively. Only gravity is present in both the brane and the bulk but not the other force of the standard model. 

Let us begin with the action that we have taken in 4D Einstein-Hilbert action for the bulk added:
\begin{equation}\label{eq:actionDGP}
S=M^3 \int d^5 X \sqrt{-g_{(5)}}(R_{(5)}-\mathcal{L}_{m}) +M^2_{P}\int d^4x\sqrt{-g}R,
\end{equation}
where $g_{(5)}$ and $g$ are the determinants of the metric of the five-dimensional bulk $g_{AB}^{(5)}$ and four-dimensional brane $g_{\mu\nu}$ respectively, t, and $R_{(5)}$ and $R$ are their corresponding Ricci scalars. Similarly, $\mathcal{L}_m$ is the Lagrangian associated with the fields confined on the brane, included if we consider the CC as a fluid. 
Therefore, the induced metric is defined as usual from the bulk metric as $g_{\mu\nu}=\partial_{\mu}X^{A}\partial_{\nu}X^{B}g^{(5)}_{AB}$. Notice that the capital letters runs as $A,B=0,1,2,3,4$ and greeks letters runs as $\mu,\nu=0,1,2,3$.
 
Thus, the background expansion rate in the DGP model using a flat FRW metric can be written as (see \cite{Deffayet:2000uy} for details):
\begin{eqnarray}\label{eq:evol_DGP}
H(z)^2 &=& H_0^2 \left[\sqrt{\Omega_m (1+z)^3 +\Omega_{r}(1+z)^4 +\Omega_{\Lambda}+\Omega_{rc}} 
\right. \nonumber \\ && \left.
\pm \sqrt{\Omega_{rc}}\right]^2,
\end{eqnarray}
where $H_0 =100h\text{km}/\text{s}\text{Mpc}^{-1}$ is the expansion rate today, $\Omega_m$ represents the fractional matter density today, $\Omega_\Lambda$ the CC term and $\Omega_{rc}=(4H_{0}^{2}r_{c}^{2})^{-1}$. Here, in addition to the matter and the crossover scale contributions we have included the radiation term.

We can compare (\ref{eq:evol_DGP}) with the standard flat Friedmann evolution equation with a dark energy component $\Omega_{DE}$:
\begin{eqnarray}
H^2(z) &=& H^{2}_{0} \left[\Omega_m (1+z)^3 +\Omega_r (1+z)^4 
\right. \nonumber\\ && \left.
+ \Omega_{DE} (1+z)^{3(1+\omega_{DE})}\right],
\end{eqnarray}
where $\omega_{DE}$ is the EoS for the DE component. Comparing the latter with (\ref{eq:evol_DGP}) we observe that $(\Omega_{rc} +\Omega_{\Lambda})$ behaves similarly to an effective CC.

If we set the $z=0$ value in (\ref{eq:evol_DGP}) leads to the constraint condition:
\begin{equation}\label{eq:constraintC}
\sqrt{\Omega_m +\Omega_r +\Omega_{\Lambda} +\Omega_{rc}} \pm \sqrt{\Omega_{rc}} =1,
\end{equation}
which differs from the conventional $\Omega_m +\Omega_r +\Omega_{DE} =1$. Therefore, from (\ref{eq:constraintC}) we get
\begin{equation}
\Omega_{rc} = \frac{1}{4} \left(\Omega_m +\Omega_r +\Omega_{\Lambda} -1\right)^2.
\end{equation}
The latter shows that for a flat universe with radiation component, $\Omega_{rc}$ is \textit{always} smaller that $\Omega_{DE}$. Even more, at large scales ($\Omega_{\Lambda}\approx 0.7$, $\Omega_m \approx 0.3$, $\Omega_r=2.469\times10^{-5}h^{-2}(1+0.2271\times N_{eff}$, $h=H_0/100$kms$^{-1}$Mpc$^{-1}$, and $N_{eff}=3.04$) the $\Omega_{rc}$ vanishes and we obtain the standard cosmology with a CC.

We observe from the evolution equation (\ref{eq:evol_DGP}) that there are two branches:
considering the positive sign emerges the branch in where is necessary to introduce a CC (i.e $\Omega_{\Lambda}\neq 0$) to drive a late cosmic acceleration. Considering the negative sign it is not necessary to add a CC (i.e $\Omega_{\Lambda}= 0$) component to describe acceleration at late-time. This latter is however ruled out by supernovae data \cite{fang}. 

\begin{figure}
\centering
\includegraphics[width=0.45\textwidth,origin=c,angle=0]{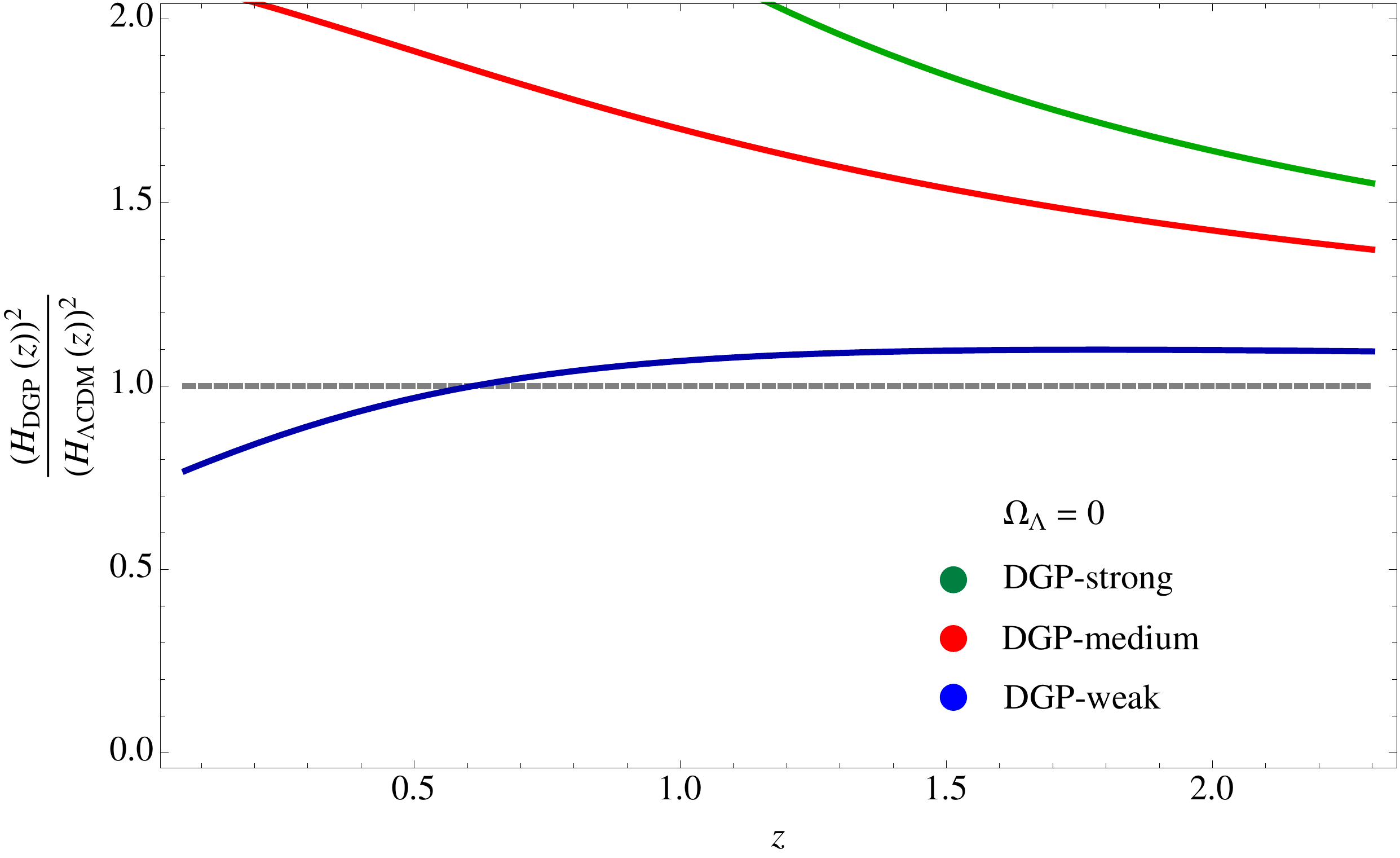}
\includegraphics[width=0.45\textwidth,origin=c,angle=0]{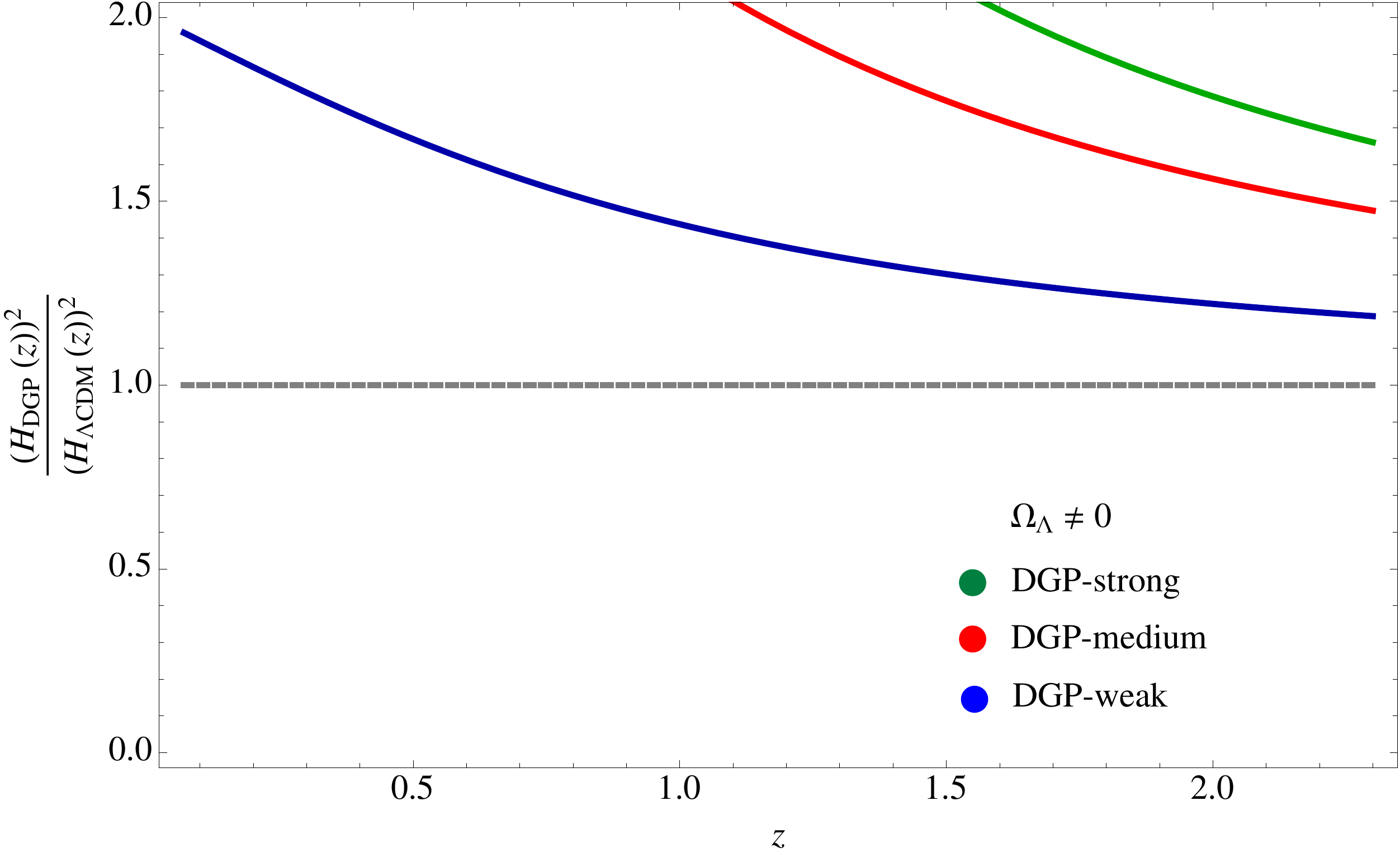}
\caption{$H(z)^2$ ratio between DGP stable model and $\Lambda$CDM model. The curves represents the cases in where the strength (weakness) DGP stable model can be fixed. \textit{Left:} Evolution of the $H(z)_{\text{DGP}}^{2}/H(z)_{\Lambda \text{CDM}}^{2}$ with a $\Omega_\Lambda =0$. \textit{Right:} Evolution of the $H(z)_{\text{DGP}}^{2}/H(z)_{\Lambda \text{CDM}}^{2}$ with a $\Omega_\Lambda \neq 0$.
} \label{fig:evolution_dgp}
\end{figure}

Therefore we consider three values of the cross-over scale: $r_c H_0 =0.2$, $r_c H_0 =0.6$ and $r_c H_0 =1.9$, which we renamed as: DGP strong, DGP medium and DGP weak stable models, respectively. The advantage of these slightly changes over the values in comparison to  \cite{Barreira:2016ovx} is that we can observe in Figure \ref{fig:evolution_dgp} a distinguishable difference between each DGP stable model and $\Lambda$CDM at early times.

\section{DGP stable cosmological analysis}
Since for both proposals of the DGP models we have cosmic acceleration, in order to perform the analysis of the DGP stable model (with positive sign in (\ref{eq:evol_DGP})) we require observational Hubble rate data. The basic assumption of this data is due that the differential age approach estimates the Hubble rate directly from the data without assuming a specific spatial geometry or any other cosmological model. These measurements has become an effective probe in cosmology comparison with SNeIa, BAO and CMB data. Following a similar methodology from \cite{Escamilla-Rivera:2015odt}, we use the cosmic chronometer (Cosmic-C) data and we complete the dataset with six measurements of $H(z)$ obtained from BAO. We summarize these data sets as:
\subsection{$H(z)$ observations}
Usually, it is has more precision to study the observational $H(z)$ data directly due that all these test use the distance scale measurement to determinate the values of the cosmological parameters, which needs the integral of $H(z)$ and therefore loses some important information of this quantity. As an independent approach of this measure we provide two samples:
\begin{enumerate}\renewcommand\labelenumi{(\theenumi)}
\item Cosmic Chronometers (Cosmic-C) data. This kind of sample gives a measurement of the expansion
rate without relying on the nature of the metric between the chronometer and us. We are going to employ
several data sets presented in \cite{Hsamples}. A full compilation of the latter, which includes 28 measurements
of $H(z)$ in the range $0.07 < z < 2.3$, are reported in \cite{Hsamples2}. The normalized {parameter $h(z)$ can be easily
determined by considering the value $H_0 = 67.31 \pm 0.96$ km s${}^{-1}$ M~pc${}^{-1}$ \cite{Ade:2015xua}.}

\item Data from BAO. Unlike the angular diameter $d_A$ measures given by the transverse BAO
scale, the $H(z)$ data can be extracted from the measurements of the line-of-sight of this BAO scale.
Because the BAO distance scale is embodied in the CMB, its measurements on DE parameters are strongest
at low redshift. The samples that we are going to consider consist of three data points from \cite{Blake:2012pj}
and three more from \cite{Gaztanaga:2008xz} measured at six redshifts in the range $0.24 < z < 0.73$.
This data set is shown in Table \ref{tab:dataBAO}.

\begin{table}
\caption{\textit{\textit{BAO}} sample data from
 \cite{Blake:2012pj,Gaztanaga:2008xz}.
}\label{tab:dataBAO}
\centering
\begin{tabular}{ccccc}
\toprule
{\textbf{z}}  &  \boldmath{$H(z)$} \textbf{[km \boldmath{$\text{s}^{-1}$}$\bf\text{M pc}$\boldmath{$^{-1}$}]}
 & \textbf{\boldmath{${\sigma_{H}}^2$}} \\
\hline
$0.24$  & $79.69$   & $2.32$  \\
$0.34$  & $83.80$   & $2.96$  \\
$0.43$  & $86.45$   & $3.27$ \\
$0.44$   & $82.6$   & $7.8$  \\
$0.6$  & $87.9$   & $6.1$  \\
$0.73$  & $97.3$   & $7.0$ \\
\hline
\end{tabular}
\end{table}

\end{enumerate}

To perform the statistical analysis we employ (\ref{eq:evol_DGP}), where 
$(\Omega_{\Lambda}, r_c)$ are the free parameters of the model. We compute the best fits of these cosmological parameters
by minimizing the quantity
\begin{eqnarray} \label{eq:min}
\chi_{\text{H(z)}}^2
=\sum^{N}_{i=1}{\frac{\left[H_{\text{theo}}(z_i ,\Omega_m ;\Omega_{\Lambda}, r_c)-H_{\text{obs}}(z_i)\right]^2}{\sigma^{2}_{H,i}}},
\end{eqnarray}
where the $\sigma^{2}_{H,i}$ are the measurements variances and $N$ is the number of the total sample, which for our purpose will be consider three combinations between datasets.

\subsection{DGP stable model cosmological tests}
\begin{figure*}[ht]
\centering
\includegraphics[width=1.\textwidth,origin=c,angle=0]{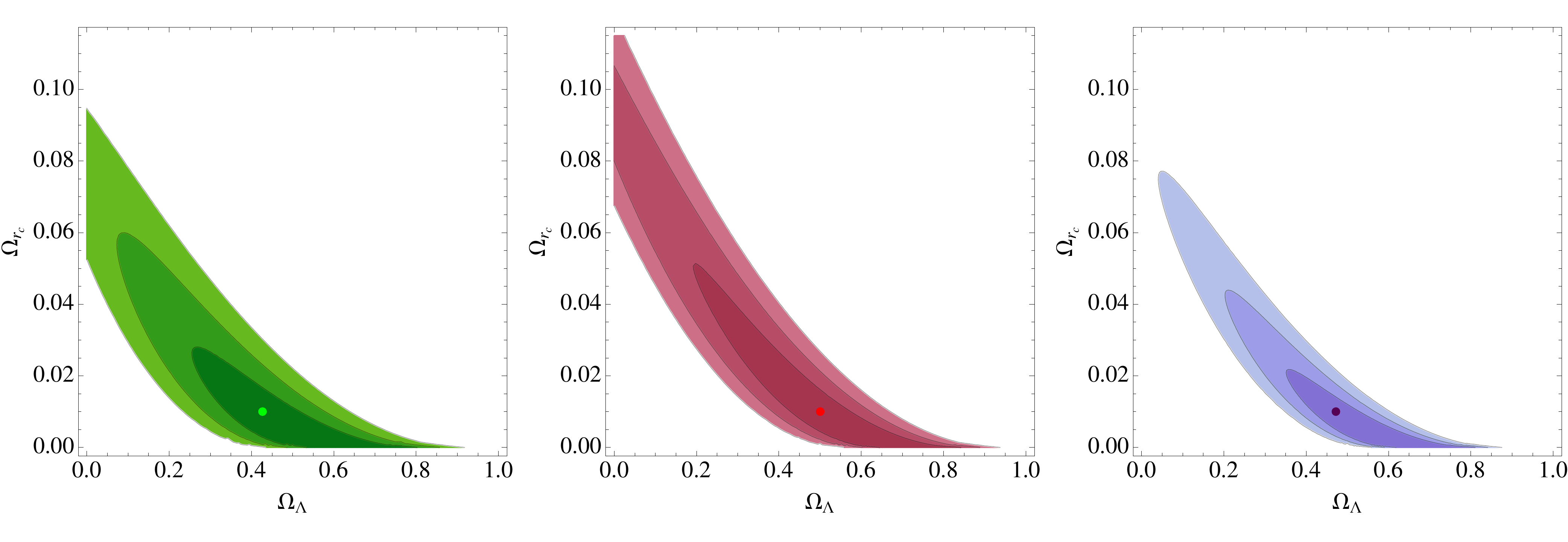}
\caption{DGP stable model confidence contours $(\Omega_\Lambda,\Omega_{rc})$ until $3$-$\sigma$. \textit{Left:} Using Cosmic-C dataset. \textit{Middle:} Using BAO dataset. \textit{Right:} Using Cosmic-C + BAO dataset. 
} \label{fig:cosmo_dgp1}
\end{figure*}

First we are going to study the case for a DGP stable model with a prior $H_0 = 67.31 \pm 0.96$ km s${}^{-1}$ M~pc${}^{-1}$ and $\Omega_m =0.315\pm 0.017$, where the set of cosmological parameters to constrains are $(\Omega_{rc},\Omega_{\Lambda})$. We perform the minimization of (\ref{eq:min}) to get the best fit values. The confidence regions in the $\Omega_m -\Omega_{rc}$ plane are show in Figure \ref{fig:cosmo_dgp1} and the statistical values are given in Table \ref{tab:datacosmo1}. 

We notice that for these priors, the cosmic acceleration at late-times is performed by the $\Omega_{\Lambda}$ term. Also, the $\Omega_{rc}$ shows a constant value for the three posible combinations of data sets.

\begin{figure*}[ht]
\centering
\includegraphics[width=1.\textwidth,origin=c,angle=0]{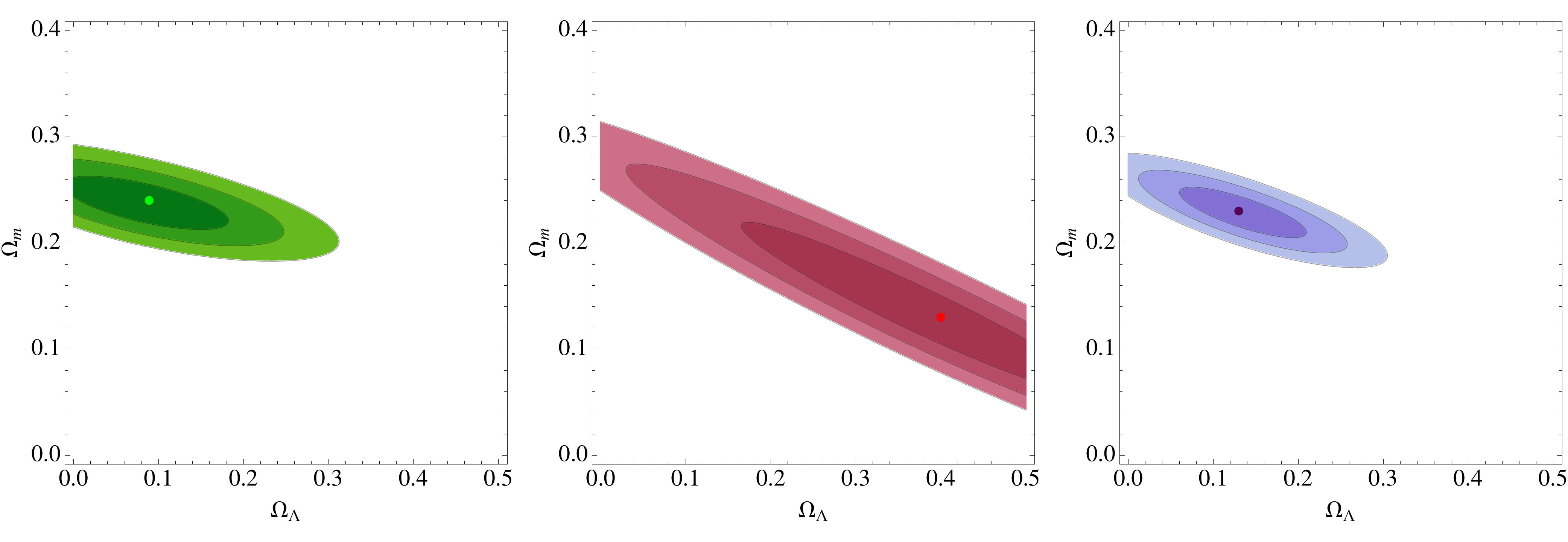}
\caption{ DGP strong model confidence contours $(\Omega_\Lambda,\Omega_m)$ until $3$-$\sigma$. \textit{Left:} Using Cosmic-C dataset. \textit{Middle:} Using BAO dataset. \textit{Right:} Using Cosmic-C + BAO dataset. 
} \label{fig:cosmo_dgp2}
\end{figure*}

For our second analysis, we consider the DGP strong stable model ($r_c H_0 =0.2$ ) with the same $H_0$ prior, where the set of cosmological parameters to constrains are $(\Omega_{m},\Omega_{\Lambda})$. The confidence regions in the $\Omega_m -\Omega_{\Lambda}$ plane are show in Figure \ref{fig:cosmo_dgp2} and the statistical values are given in Table \ref{tab:datacosmo2}. 

\begin{table}
\caption{Cosmological parameter constraints for a DGP stable model with a prior $H_0 = 67.31 \pm 0.96$ km s${}^{-1}$ M~pc${}^{-1}$ and $\Omega_m =0.31$. }
\label{tab:datacosmo1}
\centering
\begin{tabular}{ccccc}
\toprule
\textbf{Dataset}  &  $\chi^2$ &\boldmath{$\Omega_{\Lambda}$} & \boldmath{$\Omega_{rc}$} \\
\hline
Cosmo-C  & $18.827$ &$0.427\pm 0.161$   & $0.01\pm 0.177$  \\
BAO  & $4.652$ &$0.501\pm 0.235$   & $0.01\pm 0.534$  \\
Cosmo-C + BAO  & $24.404$ & $0.472 \pm 0.021$   & $0.01\pm 0.178$ \\
\hline
\end{tabular}
\end{table}

\begin{table}
\caption{Cosmological parameter constraints for a DGP strong stable model with a prior $H_0 = 67.31 \pm 0.96$ km s${}^{-1}$ M~pc${}^{-1}$ and $r_c H_0 =0.2$. }
\label{tab:datacosmo2}
\centering
\begin{tabular}{ccccc}
\toprule
\textbf{Dataset}  &  $\chi^2$ &\boldmath{$\Omega_{\Lambda}$} & \boldmath{$\Omega_{m}$} \\
\hline
Cosmo-C  & $16.984$ &$0.240\pm 0.131$   & $0.089\pm 0.221$  \\
BAO  & $3.718$ &$0.401\pm 1.635$   & $0.131\pm 1.554$  \\
Cosmo-C + BAO  & $21.329$ & $0.131 \pm 0.021$   & $0.231\pm 0.113$ \\
\hline
\end{tabular}
\end{table}
We notice in this case that $\Lambda$CDM model ($\Omega_m =0.3$ and $\Omega_\Lambda =0.7$) is discarded beyond $3$-$\sigma$. Also the best fits suggest that the cosmic acceleration in the DGP strong model is performed by only the $\Omega_{rc}$ component. 


\section{Discussion}

We notice that DGP stable model with or non addition of $\Omega_{\Lambda}$ can be distinguishable from $\Lambda$CDM at early times. Also, as we see from the Figure 1, at large redshift it seems that each DGP models starts to loiters to $\Lambda$CDM case.

Therefore, we observed some important results about the contribution of the crossover scale tested by $H(z)$ data which is shown in Figure 2, where the values for the free parameters ($\Omega_{rc}, \Omega_{\Lambda}$) are almost constant in the redshift range given by the $H(z)$ measurements. For the three confidence regions these results indicate that for our Planck priors there is no tension between these two datasets. 
In addition, the obtained value for the density parameter $\Omega_{r_c}$ is approximately equal to the value of $\Omega_{\Lambda}$, this result give us a prediction about the dominant term in the evolution equations (\ref{eq:evol_DGP}). Hence, the density of CC is the main responsible of the accelerated expansion of the universe at late times.

Indeed, the $\Lambda$CDM model is recovered for small contribution of the crossover scale density parameter.
As well, in Figure 3 we illustrate the obtained values for $\Omega_{\Lambda}$ and $\Omega_{m}$ for the DGP strong model with prior $r_c H_0 =0.2$, it is necessary to remark the difference between both values of the model. There is a tension at around 2-$\sigma$ between the two confidence contours $[\Omega_{\Lambda}-\Omega_{m}]$ using Cosmic-C and BAO. Furthermore, the obtained values from Cosmic-C, BAO and the joined dataset analysis for $\Omega_{m}$ are below the expected, the results of both densities are no consistent with the well known values for them.

Finally, we remark that for cosmological perturbations in DGP models, the main characteristics are that the integrated Sach-Wolfe (ISW) effect shows more suppression than in the standard paradigm \cite{Cardoso:2007xc} and the evolution of metric perturbations is no longer necessarily scale free \cite{Seahra:2010fj}. It is important to notice that these results could also be studied in this paper. However, to assess the impact of the brane perturbations,  a full CMB analysis should be carried out, which is beyond of the scope of this article.


\begin{acknowledgments}
G.B.-E acknowledges support from CONACYT fellowship with number 785554. C. E.-R. acknowledges support from MCTP-UNACH and M.A.G.-A. acknowledges support from SNI-M\'exico and CONACyT research fellow. Instituto Avanzado de Cosmolog\'ia (IAC) collaborations.
\end{acknowledgments}



\end{document}